# Anomalous Lattice Behavior of Vanadium Pentaoxide ($V_2O_5$): X- Ray Diffraction, Inelastic Neutron Scattering and ab-initio Lattice Dynamics

Baltej Singh[a,b], Mayanak Kumar Gupta[a], Sanjay Kumar Mishra[a], Ranjan Mittal[a,b], P. U. Sastry[a,b] Stephane Rols[c] and Samrath Lal Chaplot[a,b]

[a]*Solid State Physics Division, Bhabha Atomic Research Centre, Mumbai, 400085, India*
[b]*Homi Bhabha National Institute, Mumbai 400094, India*
[c]*Institut Laue-Langevin, BP 156, 38042 Grenoble Cedex 9, France*

We present the structural and dynamical studies of layered vanadium pentaoxide ($V_2O_5$). The temperature dependent X-ray diffraction measurements reveal highly anisotropic and anomalous thermal expansion from 12 K to 853 K. The results do not show any evidence of structural phase transition or decomposition of α-$V_2O_5$, contrary to the previous transmission electron microscopy (TEM) and electron energy loss spectroscopy (EELS) experiments. The inelastic neutron scattering measurements performed up to 673 K corroborate the result of our X-ray diffraction measurements. The analysis of the experimental data is carried out using ab-initio lattice dynamics calculation. The important role of van der-Waals dispersion and Hubbard interactions on the structure and dynamics is revealed through the ab-initio calculations. The calculated anisotropic thermal expansion behavior agrees well with temperature dependent X- ray diffraction. The mechanism of anisotropic thermal expansion and anisotropic linear compressibility is discussed in terms of calculated anisotropy in Grüneisen parameters and elastic coefficients. The calculated Gibbs free energy in various phases of $V_2O_5$ is used to understand the high pressure and temperature phase diagram of the compound. Softening of elastic constant ($C_{66}$) with pressure suggests a possibility of shear mechanism for α to β phase transformation under pressure.

# I. INTRODUCTION

The battery research is progressively increasing for the efficient use of energy to make the batteries lighter, quickly chargeable, safer, long liver, stable and cost effective[1-4]. There are relatively few gains that can be made to the battery besides directly improving component materials. The stability, safety and performance of these materials are related to their behavior (thermal expansion and phase transitions) in temperature and pressure range of interest[5-11]. The activation energy and mechanism for Li diffusion is very important to know for these materials. The accurate measurement/calculation of phonon frequencies is needed to calculate the vibrational pre-factor which is directly related to jump frequency and hence the diffusion coefficient[12]. These demanding improvements can be achieved by having an intensive microscopic understanding of atomic level dynamics of these materials[4, 13]. Vanadium based oxides show a variety of interesting properties like catalytic behavior[14, 15], metal insulator transition[16-18] and super capacitor[11, 19, 20] properties. The low cost, high abundance, easy synthesis and high energy density[21, 22] of vanadium pentaoxide ($V_2O_5$) make it useful as a cathode material for Li and Na ion batteries[4, 9, 23-30]. It also finds application in solar cells[31] as window material, antireflection coatings, critical temperature sensors[32] and light detectors. The sheets made of entangled vanadium oxide nano fibers behave like actuators[33] (artificial muscles) that contract reversibly under an electrical signal due to its good mechanical properties.

At ambient condition, vanadium pentaoxide crystallizes in orthorhombic phase (space group :*Pmmn*), known as α-$V_2O_5$[34]. The structure (Fig. 1(a) & (b)) is build-up of zig-zag double chains of $VO_5$ square-based pyramids running along b- axis. These pyramids point alternately up and down and share an edge along the chain direction [010]. The chains are bridged, through corner sharing oxygen atoms along b-axis, to make a 2-D layered structure in x-y plane. The layers are weakly bonded along c- axis via V-O van der Walls interaction. There are three different types of O atoms named as: vanadyl(O2), bonded to a single vanadium that forms the apex of the pyramids; chain (O3) that binds to three vanadium; and bridge (O1) that is bonded to two vanadium atoms and couples the chains together. The V atom inside the pyramid is shifted towards the apex O2 atom.

Vanadium pentoxide has been studied by a variety of experimental techniques over a range of temperatures. Thermal expansion coefficients of α-$V_2O_5$ as measured using dilatometer methods[35] in temperature range of 303- 723K is found to be 0.63 ×$10^{-6}$ $K^{-1}$. On the other hand the value from X- ray powder diffraction[36] in temperature range 303- 902K, is found to be 17.2 ±1.4 ×$10^{-6}$ $K^{-1}$. The thermal expansion behaviour is highly anisotropic with $α_a$ = (9.5± 0.9) x$10^{-6}$ $K^{-1}$, $α_b$ = (6.9± 1.3) x$10^{-6}$ $K^{-1}$ and $\alpha_c = (35.2 \pm 1.8) \times 10^{-6}$ K$^{-1}$. The large value of $\alpha_c$ may be due to weak van der Waal interaction between the layers along the c-axis [36]. There is a large discrepancy between the dilatometer and X-ray diffraction results of thermal expansion behaviour.

The transmission electron microscopy[37] (TEM) and electron energy loss spectroscopy (EELS)[38] experiments indicate that α-V$_2$O$_5$ undergoes decomposition above 673 K in high vacuum environment. This decomposition indicate a number of new phases, which may be related to ordered super-lattices of anion (O$^{-2}$) vacancies in vanadium pentoxide with no evidence of crystallographic shear[37]. The EELS measurements carried out up to 873K indicated transformation of V$_2$O$_5$ to VO$_2$ & V$_2$O$_3$, which was found to be different from electron beam induced reduction studies. The initial thermal decomposition of V$_2$O$_5$ to V$_2$O$_3$ is followed by a combination of diffusion, coalescence and stabilization process[38]. Moreover, high temperature X-ray diffraction[39] measurements carried out on xerogels of vanadium pentoxide (V$_2$O$_5$.1.6H$_2$O) indicate that it exhibit negative thermal expansion (NTE) behaviour along c- axis. This is a reversible contraction occurring through progressive dehydration with coefficient of NTE as large as $-1.5 \times 10^{-3}$ K$^{-1}$.

Vanadium pentoxide shows a rich phase diagram as a function of pressure. It exhibits variety of phase transitions[40, 41] from α-V$_2$O$_5$ to β-V$_2$O$_5$ and δ-V$_2$O$_5$ (Fig 1). A metastable phase, γ☐-V$_2$O$_5$, is also reported by deintercalating Li from γ- LiV$_2$O$_5$[42]. β- V$_2$O$_5$ is found to be the first high pressure positive electrode material in secondary Li batteries which can reversibly intercalate two Li ions. Its electrochemical behavior is quite a bit comparable to its ambient phase (α- V$_2$O$_5$) despite of being denser [43]. Balog et al report that β and δ phases could be quenched to ambient pressure and temperature.

The zone-centre phonon spectrum of α-V$_2$O$_5$ has been studied by spectroscopic experiments[44-50] and calculations based on density functional perturbation methods[51-54]. The calculated stretching modes frequencies in these studies are found to be either overestimated or underestimated depending upon the type of exchange correlation used. A comparative study[51] among bulk and monolayer structures has been made treating vanadium semi-core states 3s and 3p as bands. The relative intensities showed somewhat larger discrepancies compared to experimental Raman spectra. Moreover, high energy mode frequencies are highly overestimated[51].

Here we report our investigations of the structure and dynamics of V$_2$O$_5$ through X-ray diffraction, inelastic neuron scattering and ab-initio calculations. The main motive for the high temperature studies is to obtain unambiguous temperature dependence of lattice parameters and to study the high temperature decomposition reported previously in the literature[37, 38]. These studies provide the clarity about the reported[35, 36] discrepancies in the thermal expansion behavior of this compound. The ab-initio density functional theory studies are done with different pseudo-potentials and exchange correlation functional including the van der Waals (vdW) dispersion interaction. The optimized structure is used to calculate the phonon spectra by including the Hubbard onsite interaction (U) for the d- electron of vanadium. The neutron inelastic scattering measurements validate the phonon calculations including vdW + U interactions. The areal negative thermal expansion at low temperatures and negative linear compressibility are discussed in relation to anisotropic Grüensien parameters and elastic constants. We have also performed the calculations for the high pressure phases of $V_2O_5$. Further, the calculated Gibbs free energy for various phases of $V_2O_5$ and elastic constants as a function of pressure are used to calculate the pressure- temperature phase diagram and to understand the mechanism of the high pressure transitions in this compound.

## II. Experimental and Computational Details

X-ray diffraction studies from 12 to 853K are carried out using 18KW rotating Cu anode based powder diffractometer operating in the Bragg- Brentano focusing geometry with a curved crystal monochromator. Data were collected in the continuous scan mode at a scan speed of 1 degree per minute and step interval of 0.02 degree in the 2θ range of 10°–120°. The structural refinements were performed using the Rietveld refinement program FULLPROF[55].

The inelastic neutron scattering measurements are carried out using time of flight IN4C spectrometer at the Institut Laue Langevin (ILL), France. Thermal neutrons of wavelength 2.4 Å (14.2 meV) in neutron energy gain mode are used for the measurements. The momentum transfer, Q, extends up to 7 Å$^{-1}$. In the incoherent one-phonon approximation, the measured scattering function S (Q, E), as observed in the neutron experiments, is related to the phonon density of states[56, 57] $g^{(n)}(E)$ as follows:

$$g^{(n)}(E) = A < \frac{e^{2W_k(Q)}}{Q^2} \frac{E}{n(E,T) + \frac{1}{2} \pm \frac{1}{2}} S(Q,E) > \quad (1)$$

$$g^n(E) = B \sum_k \{\frac{4\pi b_k^2}{m_k}\} g_k(E) \quad (2)$$

Where the + or − signs correspond to energy loss or gain of the neutrons respectively and where $n(E,T) = [\exp(E/k_B T) - 1]^{-1}$. A and B are normalization constants and $b_k$, $m_k$, and $g_k(E)$ are, respectively, the neutron scattering length, mass, and partial density of states of the $k^{th}$ atom in the unit cell. The quantity between <> represents suitable average over all Q values at a given energy. 2W(Q) is the Debye-Waller factor. The weighting factors $\frac{4\pi b_k^2}{m_k}$ for various atoms in the units of barns/amu are: V: 0.1001 and O: 0.2645. The values of the neutron scattering lengths are taken from Ref.[58]. The multi-phonon contribution has been calculated using the Sjolander[59] formalism and subtracted from the experimental data.

The Vienna based ab-initio simulation package (VASP) [60, 61] is used to carry out the total energy calculation based on plane-wave pseudopotential methods. The calculations are performed using projected augmented wave (PAW) formalism of Kohn- Sham density functional theory with generalized gradient approximation (GGA) for exchange correlation as given by Perdew, Becke and Ernzerhof[62, 63]. K-point sampling was performed using 2×8×6 mesh Monkhorst-pack scheme[64] with a plane wave energy cutoff of 960 eV. The structure is optimized with atomic psuedopotentialwiths$^2$p$^4$ valence electrons and p$^6$d$^4$s$^1$ valence electrons in O and V atom respectively. Different schemes, to undertake the effect of vander Walls interaction, are available in VASP. DFT-D2 method of Grimme[65] was used, to account the van der Walls dispersive interaction. The values of C and R parameters are taken to be $C_6$=0.70, 10.80 Joule.nm$^6$.mol$^{-1}$ and $R_0$=1.342 Å, 1.562 Å for O and V atoms respectively. The on-site Coulomb interaction was accounted within the Dudarev approach[66] using $U_{eff}$ = U − J = 4.0 eV. The phonon frequencies in entire Brillouin zone are calculated using finite displacement method implemented in PHONON5.2 [67]. Hellman-Feynman forces are calculated by the finite displacement of 0.03 Å. Thermal expansion calculation is done using pressure dependence of phonon frequencies in entire Brillouin zone in quasi harmonic approximation. Free energy calculations are done at various pressures for the completely relaxed structure, which is used to understand the high pressure phase transition in the compound.

## III. RESULTS AND DISCUSSION

### A. X-Ray Diffraction

The powder X-ray diffraction measurements are performed (Fig 2)in the temperature range 12-853 K that extends beyond the reported decomposition[37] temperature 673 K.The measurements are used to investigate the phase stability and thermal expansion behaviour of the compound. Fig 2 shows the evolution of powder diffraction pattern at selected temperatures. For clarity the diffraction patterns are shown (Fig 2) only upto 2θ=35°. It is evident from the figure that we do not observe any dramatic change in the powder diffraction pattern with temperature. We find that on increasing temperature from 12 K, the most intense peak in the diffraction pattern around (2θ=20.3°), corresponding to (001) Bragg reflection, shifts toward lower 2θ value. This peak corresponds to the interlayer distance of distorted square pyramids ($VO_5$) lying perpendicular to the orthorhombic c-axis. The shift is due to the expansion of the lattice along the c- axis (Fig 2), with increase of temperature. Moreover, the Bragg peaks (201) and (301) around 2θ= 25.7 and 31.1 degree respectively, also shift towards lower 2θ that imply an expansion in the a-c plane but with different rate. The Braggs profile at 31.1 degree corresponds to two peaks, (400) and (301), and show clear splitting above 573 K which is due to prominent increase in *c*-lattice parameter as compared to that of the *a*-lattice parameter. In the diffraction pattern a very low intensity peak appears above 500K at around 2θ=28° marked with arrow in the Fig 2. The diffraction pattern was compared with the reported[37] decomposed phases of $V_2O_5$; however, this did not signify the decomposition of $V_2O_5$. All the peaks present in the diffraction patterns can be indexed with orthorhombic structure with *Pmmn* space group and consistent with literature[34]. The refined lattice parameters, bond lengths and bond angles are given in TABLE I.

The variation of structural parameters with temperature is shown in Fig 2. On heating above 12 K, the *a* lattice parameter decreases up-to 250 K and then increases above 250 K. The *c* parameter increases in the entire temperature range; however, the *b* parameter does not show significant temperature dependence. It is evident from this study that the thermal expansion behaviour is highly anisotropic with $\alpha_a = (-1.9 \pm 0.1) \times 10^{-6} K^{-1}$ for temperature range 12- 250 K and $\alpha_a = (3.3 \pm 0.1) \times 10^{-6} K^{-1}$ for above 250 K. Thermal expansion coefficient along *b* and *c* are $\alpha_b = (1.9 \pm 0.2) \times 10^{-6} K^{-1}$ for temperature below 413K and $\alpha_b = (-1.7 \pm 0.2) \times 10^{-6} K^{-1}$ above 413K and $\alpha_c = (42.2 \pm 0.5) \times 10^{-6} K^{-1}$ respectively. The large value of $\alpha_c$ may be due to weak van der Waal interaction between the layers along the c-axis. As a result, the $VO_5$ polyhedral layers move further apart from each other with increase of temperature. On the average a 3.7 % change in volume is observed on increase of temperature from 12K up to 853 K. We did not observe any signature of decomposition of $V_2O_5$ at around 673 K as reported previously[37] using transmission electron microscopy.

**B. Role of van-der Walls and Hubbard onsite Interaction in $V_2O_5$**

α- $V_2O_5$ is a layered compound and layered structures are generally bonded by weak van der Waals (vdW) interactions[68, 69]. Moreover, V atom has strongly correlated *d* electrons in its valence shell that contribute in bonding[70], which implies that the Hubbard onsite interaction (U) can play an important role in this compound. Earlier density functional theory calculations without the above interactions were performed for partially relaxed (at fixed volume) structure of $V_2O_5$ keeping lattice parameters fixed to experimental values[51-53, 71-73]. Other first principles calculations on $V_2O_5$ highly overestimated the c- lattice parameters, and hence produced large discrepancies in physical properties[41, 51-53, 72, 74-79, 17, 65]. Recent studies[80-86] considered the van der wall interaction that shows a good agreement of the c lattice parameter with experiments. However, detailed investigation about the role of the van der Waals interaction on bonding, phonon spectra and other thermo-dynamical properties of $V_2O_5$ is not available. Since all the phases of $V_2O_5$ possess layered structures[41, 43, 72], it is expected that the van der Waals dispersion interactions would also play an important role in structural stability of high pressures phases of $V_2O_5$. Being an important material for a wide range of applications, it is important to study various thermodynamical properties of $V_2O_5$ over a range of pressures and temperatures.

In order to bring out the importance of the van der Waals(vdW) and the onsite Hubbard(U) interactions, various optimized structures of α-$V_2O_5$ with only GGA, GGA+U, GGA+vdW and GGA+vdW+U respectively are compared with the experimental values obtained from our X- ray diffraction experiments [TABLE I]. The structural data as obtained from GGA approximation gives highly overestimated (≈10%) volume in comparison to the experimental value. The large deviation from the experimental lattice parameters, in particular the c parameter, is found when the van der Walls dispersion interactions are not considered. A parameter based van der Wall approach implemented along with the GGA exchange correlation excellently reproduces the experimental lattice parameters (TABLE I). Moreover, the calculated V-O bond lengths and bond angles by this approach agree very well with the experimental values.

The charge density in the *ac* plane as obtained from GGA calculations is plotted in Fig 3. It can be seen that V-O2 bond is a very strong bond (along c-axis) with large electron density in between the bonded atoms. The calculated charge density (Fig 3 (a)) implies that V-O1 is stronger than V-O3 bond and is consistent with the experimental bond lengths [TABLE I]. The inclusion of the van der Waals interaction (Fig 3(b)) brings $VO_5$ layers closer along 'c' axis and also modifies [TABLE I] the bond lengths and bond angles (∠V-O1-V &∠O1-V-O3). The effect of vdW interaction is found to strengthen the V-O1 bond; however, it slightly weakens the V-O3 bond [TABLE I]. The decrease in V-O1 bond length may be due to the reason that the O1 atoms is shared between two long chains of $VO_5$ polyhedral and as a result of vdW interaction these chains come close in the similar way as that of VO5 layers along c- axis. The weakest V-O3 bond expands significantly along a-axis. $O_1$ atom is covalently bonded to two V atoms along a-axis and does not allow expansion of the V-O1 bond. Moreover O2 of next layer push the O1 atom of present layer in the ab plane, increasing electron density in V-O1 bond region; strengthen the bond (Fig 3(b)). Due to the expansion of V-O3 bond lengths, *a*- lattice parameter increases slightly on inclusion of van der Walls interactions.

Further, the inclusion of the Hubbard onsite interaction [TABLE I & Fig 3(c)] for V atom results in localization of valence d electrons. This will modify the angular alignment of valence d- orbital of V atom. This strengthens the $V-O_3$ bond but weaken the $V-O_1$ bond. Charge density between the atoms (Fig 3(c)) and bond lengths [TABLE I] clearly show this feature which is consistent with the calculated Born effective charges [TABLE II]. Thus, van der Waals interaction along with Hubbard interactions plays a very important role in optimization of layered structure of α-$V_2O_5$ in line with earlier studies[52, 72, 78]. This optimized structure is further used for the calculation of other thermo-dynamical properties of α-$V_2O_5$.

These calculations are further extended to elaborate the role of van der Waals dispersion interactions among layered high pressure phases (β- and δ- $V_2O_5$) of $V_2O_5$ (Structural details of high pressure phases are given in Section III E). The structure of β- $V_2O_5$ has layers (in y-z plane) of edge and corner shared $VO_6$ octahedra, twice thick as that of alpha phase, along a- axis [Fig 1(c)]. The inclusion of van der Waals interactions correctly reproduces (TABLE III) the experimental structure in the β- $V_2O_5$, which was otherwise found to result in highly overestimated (≈14%) a- axis. The octahedral units are found to be strongly distorted with V-O distances varying from 1.569 to 2.375 Å. The effect of van der Waals interactions is more pronounced in β-phase as compared to that in α -$V_2O_5$. This is due to the geometry of these structures. The polyhedral layers in β- $V_2O_5$ form a key-lock geometry where polyhedral units in one layer are just above the voids in the next layer (Fig 1(c)). This gives rise to more numbers of O-O, V-O and V-V pairs between the polyhedral layers in β-$V_2O_5$ as compared to that in α-$V_2O_5$ and hence increases the magnitude of van der Waals interactions in β-$V_2O_5$. The lattice dynamics calculations of β-$V_2O_5$ are available in literature[72]. The earlier calculations[72] without the van der Waals interaction were performed for partially relaxed structure (at fixed volume) due to the problem of large overestimations of *a* cell parameter in the fully relaxed structure. These calculations highly overestimated the zone center phonon frequencies. The inclusion of van der Waals interaction can play a vital role in finding the correct phonon frequencies and to improve other calculated thermodynamical properties of β-$V_2O_5$.

In case of δ-$V_2O_5$, the strings of pairs of edge sharing distorted octahedra are connected through corner O atom in a zigzag fashion (Fig 1 (d)). The layers of octahedra in the a-c plane repeat along b–axis. Since the layers of $VO_6$ octahedra along b-axis are actually connected by sharing the common O atom with strong bond, so van der Waals interactions hardly play any role in the stabilization of structure of δ-$V_2O_5$. The DFT calculated structure parameters along with experimental values are given in TABLE IV. The distortion of the octahedral unit comes from V-O distances which vary from 1.6567 to 2.2227 Å. The structure is more compact with density of 4.16 g/cm$^3$ in comparison to that of β-$V_2O_5$ (3.77 g/cm$^3$) and α-$V_2O_5$ (3.37 g/cm$^3$). We found that van der Walls interaction plays an important role in stabilizing the structure of α and β phases of $V_2O_5$ and reproduces the experimentally measured lattice parameters.

## C. Phonon Spectra of α-$V_2O_5$

The temperature dependent neutron inelastic scattering measurements of α-$V_2O_5$ from 313 K up to 673 K are shown in Fig 4. As mentioned above the phonon measurements are carried out in neutron energy gain mode. So at 313 K, the low population of high energy phonons restricted us to obtain the phonon spectra only up to 110 meV. As we increase the temperature, the high energy phonons get populated and statistics of spectra at high energy improves. We did not find any significant change in the spectra at higher temperature, which is also corroborated from result of our X-ray diffraction measurements that there is no phase transition or decomposition in the temperature range of our study. The measured spectra are further used to validate the ab-initio density functional theory calculations. The calculated phonon spectra using ab-inito calculation are shown in Fig 4. The computed phonon spectrum under GGA+vdW+U is found to be in excellent agreement with the measured spectra. There exists a band gap in the phonon spectra from 110-120meV, which is also observed in previously reported IR and Raman measurements[46-48].

The calculated partial phonon densities of states of various atoms are shown in Fig 5. These calculations are useful for understanding the dynamical contribution of various atoms to the total calculated phonon density of states. There are three different types of O atoms: Vanadyl (O2) is bonded to a single vanadium and forms the apex of the pyramids; chain O3 binds to three vanadium (two along the chain direction and one in the adjacent chain), and bridge O1 is bonded to two vanadium atoms and couples the chains together. It can be seen that V-O1 and V-O3 polyhedral stretching modes are lower in energy than the V-O2 stretching mode. This behavior confirms the relatively weaker bonding of O1 and O3 with V in comparison to that of O2, which is also obvious from the bond length data [TABLE I]. Various O (O1, O2 and O3) atoms are found to contribute (Fig 5) in different energy ranges in the spectrum, which may be related to difference in V-O bond lengths.

The calculated frequencies of zone centre phonon modes are compared in TABLE V with earlier spectroscopic[46-48] and DFPT[51-53] results. The energy of modes as obtained from the previous calculations in the literature are highly overestimated in comparison to the experimental values, especially the high energy modes around 1000 cm$^{-1}$. We calculated the phonon spectra in all the schemes discussed above namely GGA, GGA+U, GGA+vdW and GGA+vdW+U. It can be seen that the presence of van der Walls dispersion interactions along with the Hubbard onsite potential significantly improved (TABLE V) the results. Especially the agreement between the experimental and calculations is much improved for the phonon modes below 350 cm$^{-1}$ and those above 900 cm$^{-1}$.

The displacement patterns of few of these modes have been plotted in Fig 6. The low energy mode at 73 cm$^{-1}$ (Fig 6) involves in-phase rigid rotation of connected VO$_5$ polyhedral units in the *ac*-plane, about the connecting O1. O1 atom is stationary in this mode. The energy of the mode is found to improve by approximately 25% (from 57 cm$^{-1}$ to 71 cm$^{-1}$) by inclusion of the van der Waals dispersion interactions in the calculations. Van der Waals interaction brings the polyhedral units, in different layers, close to each other and will enhance the vibrational energy. Another low energy mode (103 cm$^{-1}$) involves the anti-phase rotation of interconnected VO$_5$ polyhedra in *ac*-plane about O1. O1 atom also vibrates along c-axis. This mode involves the anti-phase vibrational motion along c-axis. As the calculated c lattice parameter is highly improved by van der Waals interaction, this mode frequency is improved by approximately 15%.

The high energy modes (985, 986, 1004 & 1010 cm$^{-1}$) as shown in Fig 6, involve stretching of V-O2 polyhedral bonds along the interlayer spacing. The amplitude of vibration of other atoms is negligibly small in these phonon modes. As interlayer spacing is found to be correctly reproduced by inclusion of the van der Waals and Hubbard interaction, the calculated energies of these modes are in good agreement with the experimental data. We note that the vibrational modes calculated at 856 cm$^{-1}$ is found to be highly underestimated. The mode involves the vibrations of O1 atoms in the *ab-* plane while O3 atoms are at rest. The O1 atoms in successive layers vibrate anti-parallel to each other.

**D. Elastic and Anomalous Thermal Expansion Behavior of α-V$_2$O$_5$**

The elastic constants are calculated using the symmetry-general least square method[87] as implemented in VASP5.2. The values are derived from the strain−stress relationships obtained from six finite distortions of the equilibrium lattice. For small deformations we remain in the elastic domain of the solid and a quadratic dependence of the total energy with respect to the strain is expected (Hooke's law). Plane wave energy cutoff was sufficient to converge the stress tensor. The nine independent components of elastic constant (in GPa) for the orthorhombic[88] α- V$_2$O$_5$ are calculated as:

$$\begin{pmatrix} 183.2 & 116.4 & 37.0 & 0 & 0 & 0 \\ 116.4 & 234.6 & 26.7 & 0 & 0 & 0 \\ 37.0 & 26.7 & 34.6 & 0 & 0 & 0 \\ 0 & 0 & 0 & 22.5 & 0 & 0 \\ 0 & 0 & 0 & 0 & 29.7 & 0 \\ 0 & 0 & 0 & 0 & 0 & 42.9 \end{pmatrix}$$

The very high values of $C_{11}$ and $C_{22}$ relative to $C_{33}$ imply that the structure is less compressible along *a-* and *b-* crystallographic axes. $C_{33}$ component is very small that results in high compressibility of the structure along c-axis, which is due to the layered nature of the structure along c-axis.

The quasiharmonic approximation is used to calculate the linear thermal expansion coefficients along the 'a', 'b' and 'c' -axes. The anisotropic pressure dependence of phonon energies in the entire Brillouin zone is needed for this purpose. These calculations are subsequently used to obtain the mode Grüneisen parameters. Depending on the context, the Grüneisen parameters are often defined and used in different ways in the literature[89-91]. We calculated the mode Grüneisen parameter for each phonon mode. Experimentally these are measured from pressure dependence of phonon frequencies. Theoretically, these are derived from phonon frequency change with volume (isotropic system) or lattice parameters (anisotropic system). The quasi-harmonic calculations of thermal expansion behavior are performed from the phonon frequencies at fixed absolute zero temperature. According to latest advance in Grüneisen's original work for anisotropic system[90, 91], the Grüneisen parameter is given by a general expression as

$$\Gamma_l(E_{q,i}) = -\left(\frac{\partial \ln E_{q,i}}{\partial \ln l}\right)_{T,l'} ; \; l, l' = a, b, c \; \& \; l \neq l' \tag{3}$$

Where $E_{q,i}$ is the energy of $i^{th}$ phonon mode at point **q** in the Brillouin zone. We have calculated the anisotropic Grüneisen parameters by applying an anisotropic stress of 0.5 GPa along one of the lattice parameters while keeping other lattice parameters constant, and vice versa. These calculations are performed at zero temperature. To avoid the strain in lattice angles, the symmetry of the unit cell is kept invariant on introduction of strain. The calculated mode Grüneisen parameters averaged over phonon of a particular energy E as a function of phonon energy along different directions are shown in Fig 7 (a). The Grüneisen parameters along *a*-and *b*- axes show significant negative values for the low-energy phonon modes upto 30 meV and may give rise to negative thermal expansion along a and b axes. The anisotropic linear thermal expansion coefficients are given by[90, 91]:

$$\alpha_l(T) = \frac{1}{V_0} \sum_{q,i} C_V(q,i,T) [s_{l1}\Gamma_a + s_{l2}\Gamma_b + s_{l3}\Gamma_c] ; \; l = a, b, c \tag{4}$$

Where $V_0$ is volume at ambient conditions, $C_V(q,i,T)$ is the specific heat at constant volume due to the $i^{th}$ phonon mode at point *q* in the Brillouin zone, $s_{ij}$ are the elements of elastic compliances matrix $S=C^{-1}$ as given below:

$$\begin{pmatrix} 0.0093 & -0.0038 & -0.0070 & 0 & 0 & 0 \\ -0.0038 & 0.0062 & -0.0007 & 0 & 0 & 0 \\ -0.0070 & -0.0007 & 0.0369 & 0 & 0 & 0 \\ 0 & 0 & 0 & 0.0444 & 0 & 0 \\ 0 & 0 & 0 & 0 & 0.0337 & 0 \\ 0 & 0 & 0 & 0 & 0 & 0.0233 \end{pmatrix}$$

This anisotropy in the thermal expansion behavior of this compound may arise from the anisotropy in Grüneisen parameters and elastic constants. The volume thermal expansion coefficient can be calculated as :

$$\alpha_V = \alpha_a + \alpha_b + \alpha_c \tag{5}$$

The calculated linear thermal expansion coefficients and the lattice parameters as a function of temperature are shown in Fig 7(b) and 7(c) respectively. The calculated temperature dependence of the lattice parameters is in good agreement (Fig 7(b)) with the experimental data. However, above 500 K, the calculated parameters deviate slightly from the experiments. This might be due to the fact that at high temperature the explicit anharmonicity of phonons plays an important role. This effect has not been considered in the thermal expansion calculation, which includes only the volume dependence of phonon energies (implicit contribution). The thermal expansion behaviour from eq. (4) can be expressed as

$$\alpha_a(T) \propto 0.0093\Gamma_a - 0.0038\Gamma_b - 0.0070\Gamma_c$$
$$\alpha_b(T) \propto -0.0038\Gamma_a + 0.0062\Gamma_b - 0.0007\Gamma_c \qquad (6)$$
$$\alpha_c(T) \propto 0.0070\Gamma_a - 0.0007\Gamma_b + 0.0369\Gamma_c$$

The thermal expansion coefficient along $a$ axis ($\alpha_a(T)$) has a negative value upto 300 K (Fig. 7 (d)) which becomes positive at higher temperatures. The calculated coefficients of anisotropic thermal expansion (Table III) are in agreement with our X-ray diffraction measurements and are quite different from the dilatometer measurements. As seen from above relations, the negative $\Gamma_a$ and positive $\Gamma_b$ & $\Gamma_c$ will enhance the NTE along $a$- axis. The $b$- axis shows a very small negative thermal expansion in the entire temperature range upto 1000 K. The NTE behavior along $b$-axis is governed by the negative value of $s_{21}$ and $\Gamma_b$. The $c$- axis shows a large positive value of thermal expansion in entire temperature range. As seen from above relation this behavior is related to large value of $\Gamma_c$. The calculated coefficients of thermal expansion are compared with our X-ray diffraction and previous literature [TABLE VI]. We find a very good agreement of our calculated and measured anisotropic thermal expansion coefficient. The previous dilatometer measurements have largely underestimated the thermal expansion while the previous X-ray results did not show the negative thermal expansion behaviour. The compound shows overall positive volume thermal expansion behavior. The anisotropic thermal expansion behavior is governed by the high anisotropy in Grüneisen parameters and elastic constants.

The low energy modes below 30 meV in this compound produce an anisotropic thermal expansion behavior with NTE along a and b axis at low temperatures and an expansion along c-axis in the entire temperature range. The displacement patterns of two of the low energy modes are shown in Fig 7 (e) & (f). The contribution to the thermal expansion coefficient, due to the mode at around 9 meV (assuming it as an Einstein mode with one degree of freedom), along various axes is: $\alpha_a$ = -1.3×10$^{-6}$ K$^{-1}$ $\alpha_b$ = 2.5×10$^{-6}$ K$^{-1}$, $\alpha_c$ = 7.4×10$^{-6}$ K$^{-1}$. The mode produces contraction along a-axis. Basically the mode involves (Fig 7 (e)) the out of phase motion of different polyhedral units in the same layer along c-axis. This favors VO$_5$ pyramids to come close along a-axis giving rise to negative thermal expansion along a-axis while favoring the expansion of c-axis. The other mode at around 20 meV ($\alpha_a$= -2.0×10$^{-6}$ K$^{-1}$, $\alpha_b$= -0.2×10$^{-6}$ K$^{-1}$, $\alpha_c$=10×10$^{-6}$ K$^{-1}$) involves (Fig 7(f)) the out of phase motion of V and O atoms in the a-c plane. It gives rise to overall contraction along b-axis.

**E. High Pressure Study of V$_2$O$_5$**

Orthorhombic, α-V$_2$O$_5$ is studied under high pressure by various experimental techniques like XRD[40, 41, 43, 92, 93], neutron powder diffraction[94], high-resolution transmission electron microscopy[94] spectroscopic[93, 95] and theoretical ab-initio density functional theory[41, 72, 96] (DFT) [25] etc. Raman spectra and density functional calculations[72] provide the identification of spectral finger prints specific to structural basic units of α-V$_2$O$_5$ and β-V$_2$O$_5$. β- V$_2$O$_5$ phase is obtained from α-V$_2$O$_5$ by the application of isotropic pressure between 4 to10 GPa at temperature between room temperature to 1023 K[86, 93, 95, 97, 98]. The structure[40, 41, 72, 92, 94, 99, 100] of β-V$_2$O$_5$ is monoclinic with space group P2$_1$/m. In another study[95], α-V$_2$O$_5$ to β-V$_2$O$_5$ phase transition is found to be an irreversible transition occurring above 7 GPa at ambient temperature and involves change in vanadium coordination number from 5 to 6. The microscopic mechanism of phase transition from α to β phase studied using density functional theory approach[96] indicated that the U$_5$ shear strain of the α phase causes this transition to happen via a gliding displacement of V$_2$O$_5$ layers. Further increase in pressure, beyond 7 GPa at ambient temperature, results in a recently discovered monoclinic δ-V$_2$O$_5$ phase. This phase was first obtained by Filonenko et al[99] between 8– 8.5 GPa and 873- 1373 K which they have labeled B- V$_2$O$_5$, later identified as δ-V$_2$O$_5$. However, pure δ-V$_2$O$_5$ was not obtained even up to 10 GPa[100]. This phase transforms reversibly to α-V$_2$O$_5$ [99]. The phase was characterized at 10.1 GPa and 673 K using in-situ X-ray measurement,. Another X-ray diffraction and Raman spectroscopy studies[40] up to 29 GPa and temperatures up to 1773 K on V$_2$O$_5$ is performed, which indicates that the structure of δ-V$_2$O$_5$ is monoclinic with space group C2/c. The δ-phase seems to be stable[40] from 10 up to 29 GPa, at temperatures higher than 473 K. Recently δ-V$_2$O$_5$ is obtained at P = 8.0-8.5 GPa, and T = 973–1373 K and structure is found to be independent of the synthesis method[92]. There is a large discrepancy in the reported pressures and temperatures for various phase transitions in V$_2$O$_5$.

Here we report study of the high pressure behavior of V$_2$O$_5$ from the *ab*-initio DFT calculations. The calculated pressure dependent lattice parameters for *α*- V$_2$O$_5$ along with experimental values are compared in Fig 8 (a). It can be seen that *α*- V$_2$O$_5$ shows an anisotropic compressibility along different crystallographic axes. The *a*- lattice parameter is found to expands while *c* contracts. This gives an effective negative linear compressibility (NLC) along *a*- axis and positive linear compressibility along *c* axis, while *b* axis is almost invariable in the range of applied pressure up to 6 GPa. The positive compressibility along c axis is related to decreasing interlayer distance (V-O2 interlayer in Fig 8(b)) along *c* axis. On the other hand, an increase in V-O2 intralayer bond (V-O2 bond in $VO_5$ polyhedral unit lies along *c*-axis) length signifies that V is pushed more towards the plane of $VO_5$ square pyramid. This gives rise to an expansion of square pyramid in *a-b* plane which can be seen from increasing V-V bond lengths (Fig 8(c)) along *a*- axis. The effect of expanding square plane of $VO_5$ square pyramids is more pronounced along *a* axis in comparison to that along *b* axis. The increase in V-V bond length along a-axis is governed by opening of the ∠V-O1-V and ∠O1-V-O3 bond angles (Fig 8(d)) with increase of pressure. This gives rise to a negative linear compressibility along *a* axis of α- $V_2O_5$. The diminishing difference between the *b* and *c* lattice implies a continuous change of the V coordination from square pyramidal towards octahedral. The V-O2 interlayer bond length in the *α*- phase, which is 2.8 Å at P=0, gradually reduces to 2.3 Å up to 6 GPa approaching the value in the β- phase.

We have investigated likely phase transitions by the minimization of Gibbs free energy, which is the basic requirement for the stability of one phase compared to the other phases, and is given by

$$G(T,P) = U(T,V) + PV - TS(T,V) \qquad (7)$$

Where *U* is the internal energy, *S* is phonon entropy and *P, V* and *T* are pressure, volume and temperature respectively. The phonon entropy is calculated from the phonon spectra over the entire Brillion zone under quasiharmonic approximation using

$$S(V) = -\frac{1}{N}\sum_{k,i} k_B \ln\left[1 - \exp\left(-\frac{h\nu_{k,i}(V)}{k_B T}\right)\right] + \frac{1}{N}\sum_{k,i} k_B \frac{h\nu_{k,i}(V)}{k_B T}\left[\exp\left(-\frac{h\nu_{k,i}(V)}{k_B T}\right) - 1\right]^{-1} \qquad (8)$$

The phonon spectra are first calculated as a function of pressure for the various phases using the GGA+vdW+U optimized structures. The calculated phonon dispersion and phonon density of states for various phases at ambient conditions are given in Fig. 9 for comparison. The significant gap in the phonon spectra at around 100-120 meV in α- $V_2O_5$ disappears as we go from α- $V_2O_5$ to δ- $V_2O_5$. The Gibbs free energy has been calculated (Fig 10) for the three phases (α- $V_2O_5$, β- $V_2O_5$ & δ- $V_2O_5$) as a function of pressure and temperature. Pressure is applied isotropically with a large enough cutoff to minimize the pulay stress. The dynamical stability of all the phases is observed at all the pressures upto 5 GPa when all the phonon frequencies are real.

We find free energy crossover (Fig 10) between α to β phases at around 1.5GPa at 1000 K. The critical pressure for α-β transition is somewhat underestimated from the experimental value[41] of 4.0 GPa. The α to β phase transformation is of first order and involves volume drop of about ~11%. The structure of the β phase is monoclinic containing $VO_6$ octahedral units. The $VO_6$ octahedra are found to be highly distorted [TABLE III]. Further comparison of the free energy of β and δ phase (Fig 10) indicates that δ phase is stable above 3 GPa and 1500 K. The β to δ transition is experimentally observed at 10.1 GPa and 673 K[41]. The β to δ phase transition is a monoclinic to monoclinic transition which occurs by doubling the number of atoms per unit cell. The δ- $V_2O_5$ is about ~10 % denser in comparison to β- phase. δ- $V_2O_5$ is also made of $VO_6$ distorted octahedral [TABLE IV].

The calculated and experimental[40] phase diagram of $V_2O_5$ as a function of temperature and pressure is given in Fig 11. The DFT calculations reproduce the experimental phase diagram qualitatively. The phase boundaries in these calculations are somewhat underestimated in comparison to the experimental values possibly due to the limitation of the calculations.

As we go from 'α' to 'δ' phase, V-O bond length among the layers start decreasing which decreases the distortion of $VO_6$ octahedral units. This can be seen from the difference of $(V-O)_{max}$ -$(V-O)_{min}$ [TABLE VII] for a $VO_6$ octahedral unit. In case of α- $V_2O_5$, V is five coordinated with oxygen atoms. For the sake of comparison α- $V_2O_5$ is also considered to be made up of $VO_6$ pseudo octahedra instead of $VO_5$ square pyramids. The phase transition basically takes place by decrease in volume under pressure. The volume per atom as a function of pressure (Fig 12) shows that α- $V_2O_5$ would transform to β- $V_2O_5$ and then to δ- $V_2O_5$ at higher pressures.

The calculated elastic constants for α- $V_2O_5$ as a function of isotropic pressure are given in Fig 13. It can be seen that all but $C_{66}$ components show normal hardening with increasing pressure. The $C_{66}$ component softens from 42.9 to 35.2 GPa as we increase pressure from 0 to 7.5 GPa. $C_{66}$ is a shear component given by[88]

$$C_{66} = \frac{\sigma_{xy}}{\varepsilon_{xy}} \qquad (9)$$

Where $\sigma_{xy}$ and $\varepsilon_{xy}$ represent the shear stress and shear strain respectively. Under ambient pressure conditions, the shear components $C_{44}$ and $C_{55}$ are smaller in comparison to $C_{66}$. As pressure increases the $C_{66}$ shear component becomes smaller, implying that deforming the structure in the ab-plane becomes easier on increase of pressure. It appears that the large compressibility along the c-axis alongwith the ease of deformation in ab- plane ($\varepsilon_{xy}$) could lead to the α- to β- phase transformation.

## IV. CONCLUSIONS

The experimental X-ray diffraction and inelastic neutron scattering studies along with extensive ab- initio lattice dynamical calculations shed light on the microscopic origin of anomalous lattice behavior of $V_2O_5$. The temperature dependent X-ray diffraction measurements from 12K to 853K do not show any evidence of structural phase transition or decomposition of α-$V_2O_5$, which clarify the previous ambiguity existing in the literature. The inelastic neutron scattering measurements performed up to 673 K corroborate the result of our X-ray diffraction measurements. The van der-Waals dispersion and Hubbard interactions are found to play an important role in structure and dynamics of layered $V_2O_5$. These interactions highly affect the phonon spectra and other thermodynamical properties of $V_2O_5$. The anisotropic thermal expansion calculated under quasiharmonic approximation agrees well with temperature dependent X- ray diffraction results and clarify the ambiguity in reported dilatometer measurements. The compound is found to show negative thermal expansion and negative linear compressibility along a- axis at low temperature. The anisotropic thermal expansion and anisotropic linear compressibility arises from the anisotropy in Grüneisen parameters and elastic constants. Free energy calculations shows two first ordered phase transition at high pressure and temperature i.e from α- to β- and then to δ-phase. The α to β- phase transition arises from the softening of elastic constant ($C_{66}$) with pressure which suggests a possibility of shear mechanism for this phase transformation.

**Acknowledgements**
S. L. Chaplot would like to thank the Department of Atomic Energy, India for the award of Raja Ramanna Fellowship.

TABLE I. Calculated (0K) and experimental (12K) lattice parameters and bond lengths (VO$_5$ Polyhedra) of α-V$_2$O$_5$. The c- axis is perpendicular to layers of VO$_5$ square pyramids. V-O2 (pseudo octahedral bond) represents the bond length of V in the one layer and oxygen in the next layer.

| | Experimental | DFT (GGA) | DFT (GGA+ U) | DFT (GGA+ vdW) | DFT (GGA+ vdW+U) |
|---|---|---|---|---|---|
| a( Å) | 11.523(2) | 11.572 | 11.564 | 11.661 | 11.649 |
| b( Å) | 3.562(6) | 3.577 | 3.638 | 3.544 | 3.608 |
| c( Å) | 4.330(1) | 4.747 | 4.728 | 4.424 | 4.408 |
| V( Å$^3$) | 177.74(8) | 196.50 | 198.88 | 182.83 | 185.25 |
| E/atom (eV) | | -8.41 | -7.26 | -8.53 | -7.38 |
| V-O1 | 1.791(8) | 1.793 | 1.809 | 1.786 | 1.802 |
| V-O2 | 1.554(8) | 1.599 | 1.609 | 1.602 | 1.614 |
| V-O3 | 1.877(0) | 1.892 | 1.916 | 1.880 | 1.904 |
| | 2.027(6) | 2.046 | 2.030 | 2.063 | 2.044 |
| | 1.877(0) | 1.892 | 1.916 | 1.880 | 1.904 |
| V-O2 (Pseudo-Octahedral bond) | 2.775(1) | 3.149 | 3.119 | 2.822 | 2.795 |
| ∠V-O1-V | 146.838(9) | 146.5 | 146.2 | 149.5 | 149.4 |
| ∠O1-V-O2 | 105.574(1) | 105.1 | 105.0 | 104.6 | 104.5 |
| ∠O2-V-O3 | 104.721(5) | 105.1 | 104.7 | 105.2 | 104.7 |
| | 104.799(2) | 107.9 | 107.9 | 105.9 | 105.9 |
| | 104.721(5) | 105.1 | 104.7 | 105.2 | 104.7 |
| ∠O1-V-O3 | 96.383(9) | 97.0 | 96.5 | 97.7 | 97.1 |
| | 149.626(7) | 147.1 | 147.0 | 149.5 | 149.6 |
| | 96.383(9) | 97.0 | 96.5 | 97.7 | 97.1 |
| ∠O3-V-O3 | 75.818(9) | 74.5 | 75.2 | 74.3 | 75.2 |
| | 143.270(1) | 141.9 | 143.3 | 141.1 | 142.8 |

TABLE II. Calculated Born effective charges for atoms present at various Wyckoff sites in α-$V_2O_5$. $Z_{I,XX}$, $Z_{I,YY}$ and $Z_{I,ZZ}$ are the diagonal components of Born charge tensor.

| Atom | Calculation | $Z_{I,xx}$ | $Z_{I,yy}$ | $Z_{I,zz}$ |
|---|---|---|---|---|
| V | GGA | 7.34 | 5.76 | 2.40 |
| | GGA+U | 6.69 | 5.24 | 2.51 |
| | GGA+vdW | 6.55 | 6.07 | 2.46 |
| | GGA+vdW+U | 6.42 | 5.78 | 2.92 |
| O1 | GGA | -6.21 | -0.87 | -0.70 |
| | GGA+U | -6.58 | -0.74 | -0.78 |
| | GGA+vdW | -7.27 | -0.85 | -0.78 |
| | GGA+vdW+U | -7.45 | -0.80 | -0.86 |
| O2 | GGA | -0.67 | -0.56 | -1.42 |
| | GGA+U | -0.55 | -0.56 | -1.52 |
| | GGA+vdW | -0.55 | -0.61 | -1.64 |
| | GGA+vdW+U | -0.55 | -0.59 | -1.92 |
| O3 | GGA | -2.89 | -4.78 | -0.53 |
| | GGA+U | -2.57 | -4.32 | -0.56 |
| | GGA+vdW | -2.45 | -5.04 | -0.44 |
| | GGA+vdW+U | -2.27 | -4.79 | -0.54 |

TABLE III. Calculated and experimental lattice parameters and bond lengths of β-$V_2O_5$. The experimental data[40] is obtained from post-pressure and temperature quenched β-phase.

| | Bonds | Experimental[40] | GGA | GGA+vdW+U |
|---|---|---|---|---|
| a( Å) | | 7.1118 | 8.4025 | 7.1803 |
| b( Å) | | 3.5791 | 3.5730 | 3.6215 |
| c( Å) | | 6.2903 | 6.3527 | 6.3218 |
| B | | 90.15 | 87.7931 | 90.0343 |
| V( Å$^3$) | | 160.11 | 190.58 | 164.39 |
| E/atom (eV) | | | -8.39 | -7.38 |
| V1-Octahedra | V1-O1 | 1.634 | 1.5904 | 1.60159 |
| | V1-O2 | 2.214 | 2.0538 | 2.00806 |
| | V1-O3 | 1.884 | 1.8914 | 1.90802 |
| | | 1.884 | 1.8914 | 1.90802 |
| | V1-O4 | 2.325 | 2.3753 | 2.34176 |
| | V1-O5 | 1.731 | 1.8801 | 1.90704 |
| V2-Octahedra | V2-O2 | 1.569 | 1.6538 | 1.67723 |
| | V2-O3 | 2.018 | 2.1683 | 2.12214 |
| | V2-O4 | 1.868 | 1.8754 | 1.89376 |
| | | 2.318 | 2.3633 | 2.31304 |
| | | 1.868 | 1.8754 | 1.89376 |
| | V2-O5 | 1.873 | 1.7434 | 1.75291 |

TABLE IV. Calculated and experimental lattice parameters and bond lengths of δ-$V_2O_5$. The experimental data[40] is obtained from post-pressure and temperature quenched δ-phase.

|  | **Experimental[40]** | **GGA** | **GGA+vdW+U** |
|---|---|---|---|
| a(Å) | 11.9719 | 11.9974 | 12.0303 |
| b(Å) | 4.7017 | 4.7041 | 4.6880 |
| c(Å) | 5.3253 | 5.4169 | 5.2991 |
| β | 104.4 | 103.27 | 103.98 |
| V(Å³) | 290.32 | 297.56 | 290.01 |
| E/atom(eV) |  | -8.36 | -7.38 |
| V-O1 | 1.811 | 1.803 | 1.815 |
| V-O2 | 2.112 | 2.078 | 2.028 |
|  | 1.628 | 1.657 | 1.684 |
| V-O3 | 1.783 | 1.788 | 1.833 |
|  | 2.017 | 2.036 | 2.016 |
|  | 2.145 | 2.223 | 2.130 |

TABLE V. Calculated and experimental frequencies of zone center phonon modes in α-phase of $V_2O_5$. The calculations in the present study are performed using the optimized minimum energy structures while the previous calculations[51-53] were performed at experimental fixed volumes.

|        | Expt. (Raman & IR) | GGA  | GGA+U | GGA+vdW | GGA+vdW+U | LDA-FHI[52] | GGA[53] | LDA-HGH[51] |
|--------|--------------------|------|-------|---------|-----------|-------------|---------|-------------|
| Au     | 76                 | 66   | 70    | 66      | 70        |             | 127     |             |
| B3u(I) | 72                 | 57   | 57    | 71      | 73        | 74          | 102     | 72          |
| Ag(R)  | 107                | 85   | 88    | 98      | 103       | 108         | 137     | 104         |
| B2g(R) | 147                | 125  | 130   | 134     | 140       | 146         | 143     | 145         |
| B1u(I) | 140                | 118  | 118   | 139     | 141       | 142         | 248     | 136         |
| B1g(R) | 147                | 139  | 152   | 150     | 162       | 167         | 169     | 147         |
| B3g(R) | 149                | 139  | 152   | 151     | 162       | 169         | 174     | 148         |
| Ag(R)  | 200                | 172  | 179   | 189     | 197       | 193         | 291     | 185         |
| B2g(R) | 201                | 181  | 187   | 194     | 201       | 200         | 259     | 192         |
| B2u(I) | 213                | 208  | 210   | 210     | 210       | 205         | 254     | 217         |
| B3g(R) | 232                | 231  | 231   | 227     | 226       | 219         | 273     | 230         |
| B3u(I) | 259                | 244  | 242   | 258     | 256       | 255         | 327     | 267         |
| Au     | 263                | 257  | 251   | 266     | 260       |             | 297     |             |
| B2u(I) | 285                | 274  | 277   | 281     | 281       | 271         | 304     | 285         |
| B1g(R) | 290                | 273  | 273   | 282     | 281       | 277         | 305     | 285         |
| B3g(R) | 291                | 274  | 280   | 285     | 289       | 279         | 311     | 284         |
| B1u(I) | 290                | 275  | 276   | 286     | 289       | 295         | 329     | 291         |
| B2g(R) | 315                | 283  | 275   | 300     | 294       | 299         | 385     | 309         |
| Ag(R)  | 310                | 286  | 279   | 301     | 295       | 294         | 383     | 303         |
| B3u(I) | 302                | 346  | 358   | 329     | 345       | 362         | 347     | 339         |
| B1u(I) | 354                | 349  | 342   | 347     | 340       | 338         | 439     | 358         |
| B2g(R) | 350                | 343  | 344   | 347     | 351       | 361         | 388     | 352         |
| Ag(R)  | 404                | 387  | 381   | 389     | 385       | 382         | 454     | 396         |
| B3u(I) | 411                | 412  | 358   | 418     | 420       | 478         | 463     | 414         |
| Ag(R)  | 483                | 466  | 381   | 452     | 468       | 528         | 484     | 467         |
| B1u(I) | 472                | 472  | 468   | 467     | 461       | 477         | 481     | 464         |
| B2g(R) | 507                | 480  | 494   | 478     | 487       | 529         | 500     | 482         |
| Au     | 533                | 473  | 493   | 492     | 510       |             | 468     |             |
| B2u(I) | 506                | 473  | 493   | 493     | 511       | 602         | 474     | 495         |
| Ag(R)  | 528                | 514  | 515   | 520     | 518       | 542         | 539     | 518         |
| B1u(I) | 570                | 514  | 519   | 525     | 534       | 665         | 623     | 617         |
| B3g(R) | 702                | 676  | 687   | 685     | 696       | 772         | 676     | 693         |
| B1g(R) | 702                | 676  | 686   | 685     | 695       | 772         | 676     | 693         |
| B3u(I) | 766                | 758  | 758   | 778     | 779       | 849         | 768     | 758         |
| B2g(R) | 963                | 840  | 831   | 861     | 856       | 1010        | 936     | 929         |
| B1u(I) | 974                | 1046 | 1006  | 1029    | 985       | 1014        | 1007    | 1014        |
| B3u(I) | 981                | 1043 | 1002  | 1030    | 986       | 1011        | 1006    | 1011        |
| B2g(R) | 992                | 1060 | 1023  | 1044    | 1004      | 1030        | 1024    | 1029        |
| Ag(R)  | 992                | 1066 | 1030  | 1050    | 1010      | 1029        | 1026    | 1031        |

TABLE VI. Calculated and experimental thermal expansion coefficients of α-$V_2O_5$.

| Thermal Expansion Coefficient ($\times 10^{-6}$) $K^{-1}$ | Dilatometer Measurements[35] (303-723K) | X-Ray Diffraction[36] (303-902K) | X-Ray Diffraction (Our Study) 12-853K | Ab-initio lattice dynamical calculations |
|---|---|---|---|---|
| $\alpha_a$ | …… | 9.5± 0.9 | -1.9± 0.1 (12-250K) <br> 3.3± 0.1 (above 250K) | -1.3 (110 K) <br> 2.9 (above 250K) |
| $\alpha_b$ | …... | 6.9± 1.3 | 1.9± 0.2 (upto 413K) <br> -1.7± 0.2 (above 413K) | -2.3 |
| $\alpha_c$ | ……. | 35.2± 1.8 | 42.2± 0.5 | 46.2 |
| $\alpha_V$ | 0.63 | 51.6 ±1.4 | 44.5± 0.5 | 47.1 |

TABLE VII. Experimental and calculated octahedral distortions in terms of bond length. For comparison, α-$V_2O_5$ is also considered to be made up of $VO_6$ octaherda instead of $VO_5$ square pyramids. The experimental parameters[40] are obtained at room temperature and ambient pressure while calculations are at 0 K and 0 GPa.

| Phase | Experimental[40] | | | Calculated | | |
|---|---|---|---|---|---|---|
| | Max. V-O | Min. V-O | Max-Min | Max. V-O | Min. V-O | Max-Min |
| α-$V_2O_5$ | 2.791 | 1.577 | 1.214 | 2.795 | 1.614 | 1.181 |
| β-$V_2O_5$ | 2.325 | 1.634 | 0.691 | 2.364 | 1.592 | 0.772 |
| δ-$V_2O_5$ | 2.145 | 1.628 | 0.517 | 2.189 | 1.657 | 0.532 |

FIG. 1 (Color online) (a) structure of α-V$_2$O$_5$, containing square pyramids (b) α-V$_2$O$_5$ containing pseudo octahedra (for comparison) (c) β-V$_2$O$_5$ containing two different types of VO$_6$ octahedra (d) δ-V$_2$O$_5$ containing VO$_6$ octahedra. The labels of the O atoms in δ-V$_2$O$_5$ are different from that of α-V$_2$O$_5$.

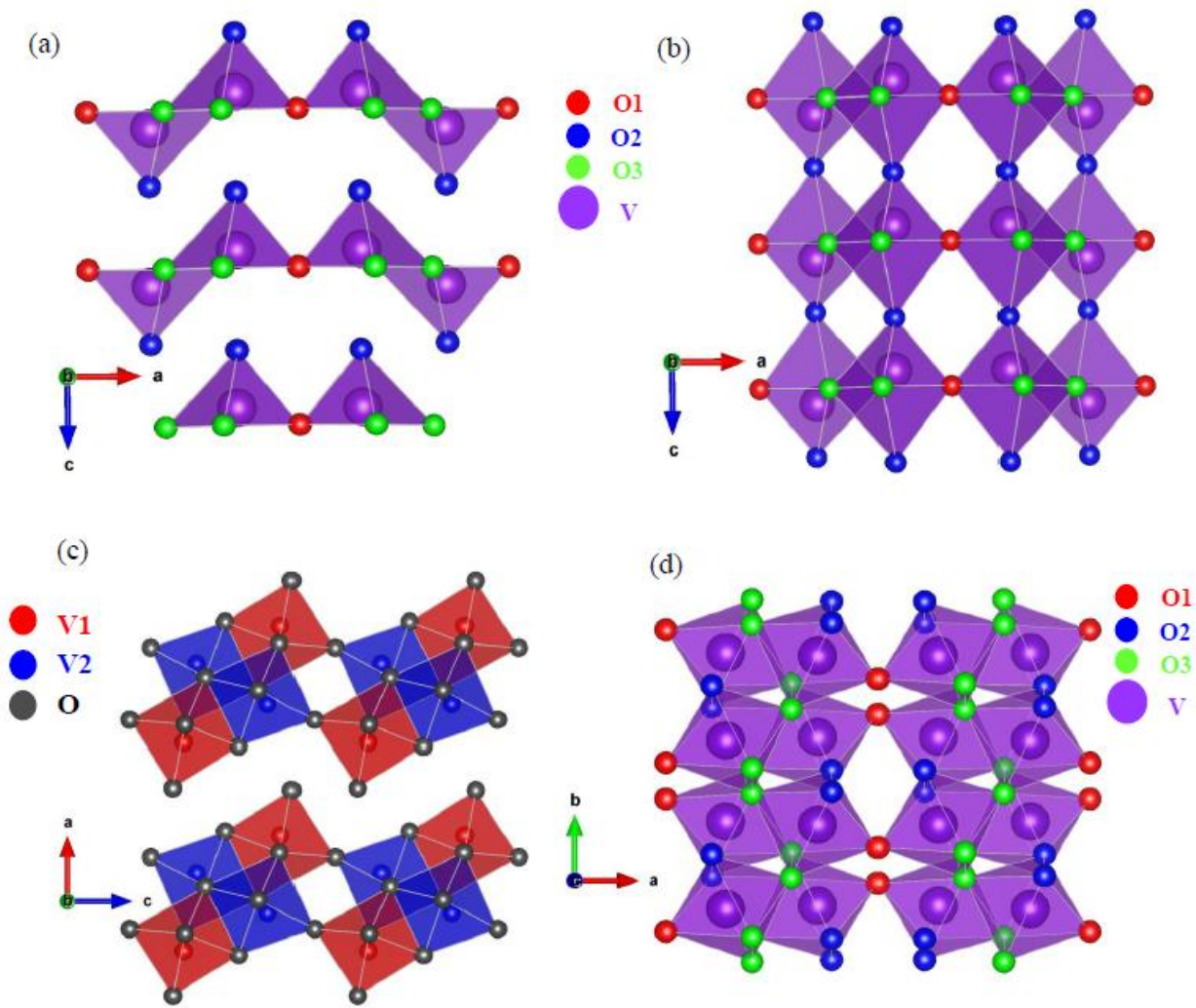

FIG. 2 (Color online) Evolution of X-Ray diffraction pattern (left panel) and structural parameters (right) with temperature.

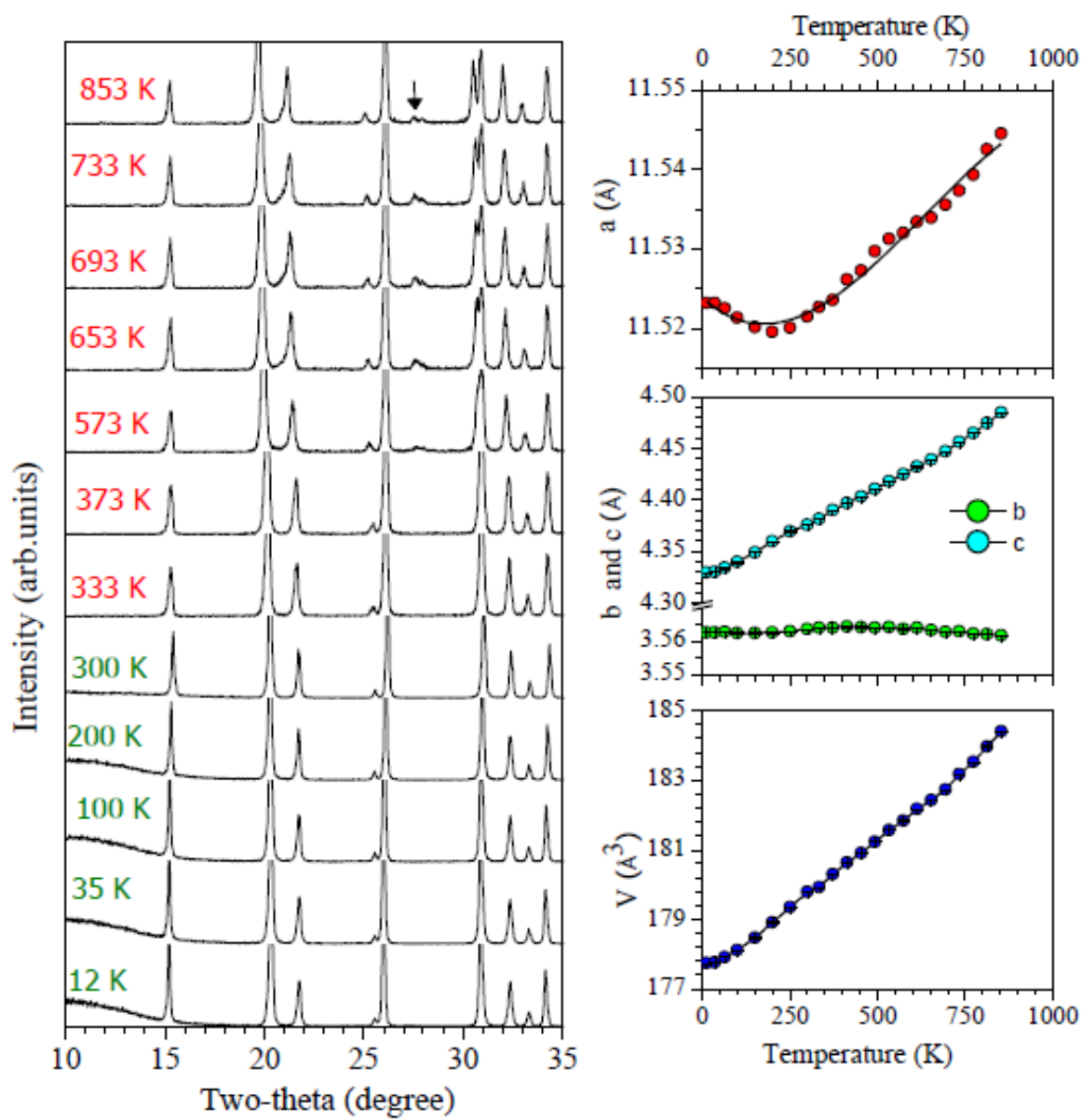

FIG. 3 (Color online) The atomic charge density plot of α-V$_2$O$_5$, in the a-c plane, for calculations performed considering (Top) GGA, (Middle) GGA+vdW, and (Bottom) GGA+vdW+U. The charge density scale is displayed on the right. The atomic positions in the space group Pmmn are O1 at 2a(x,y,z), O2 at 4f(x,y,z), O3 at 4f(x,y,z), and V at 4f(x,y,z) Wyckoff site.

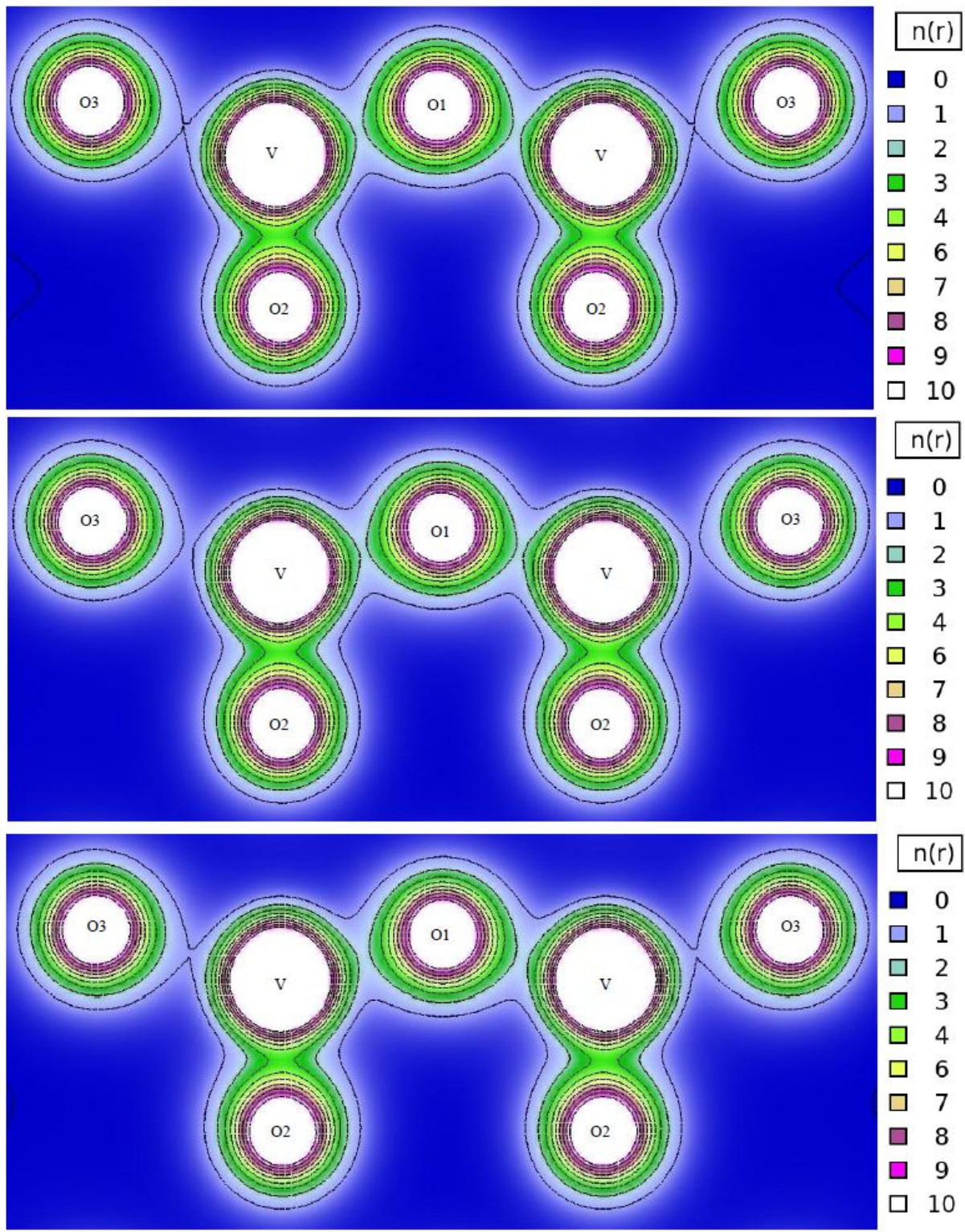

FIG.4 (Color online) (a) Experimental and calculated (GGA+vdW+U) neutron-weighted phonon density of states in α-$V_2O_5$. (b) Phonon density of states calculated using various types of interactions in α-$V_2O_5$.

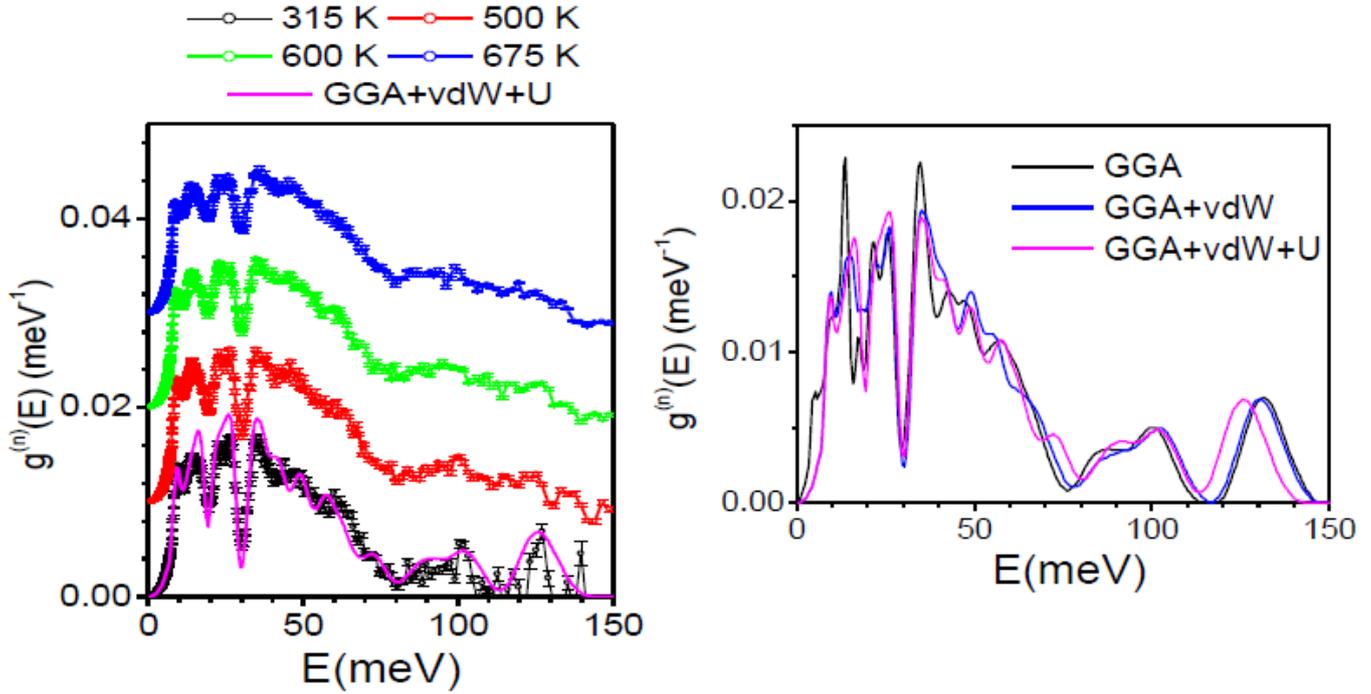

FIG.5 (Color online) Calculated (GGA+vdW+U) partial phonon density of states of atoms at various Wyckoff sites in α-$V_2O_5$. The atomic positions in the space group Pmmn are O1 at 2a(x,y,z), O2 at 4f(x,y,z), O3 at 4f(x,y,z), and V at 4f(x,y,z) Wyckoff site.

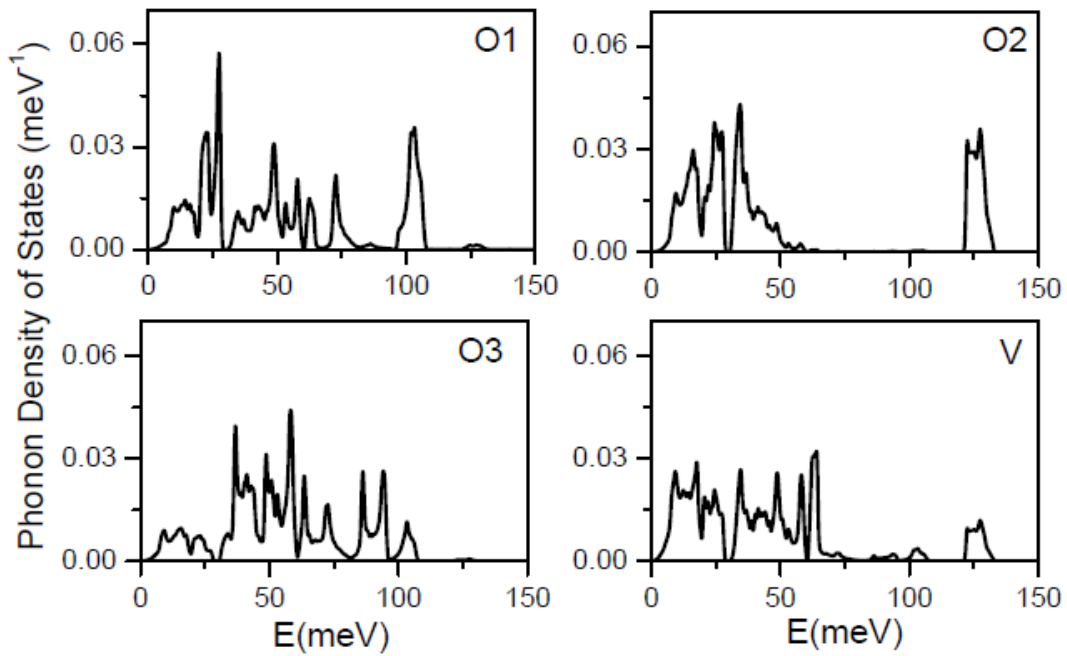

FIG.6 (Color online) The calculated eigen vectors and frequencies (in cm$^{-1}$) of zone centre modes that are highly affected by van der Waals and Hubbard onsite interaction in α- $V_2O_5$. The values of the frequencies correspond to our GGA+vdW+U calculations. key: V(Violet),O1(Red), O2(Blue) ,O3(Green).

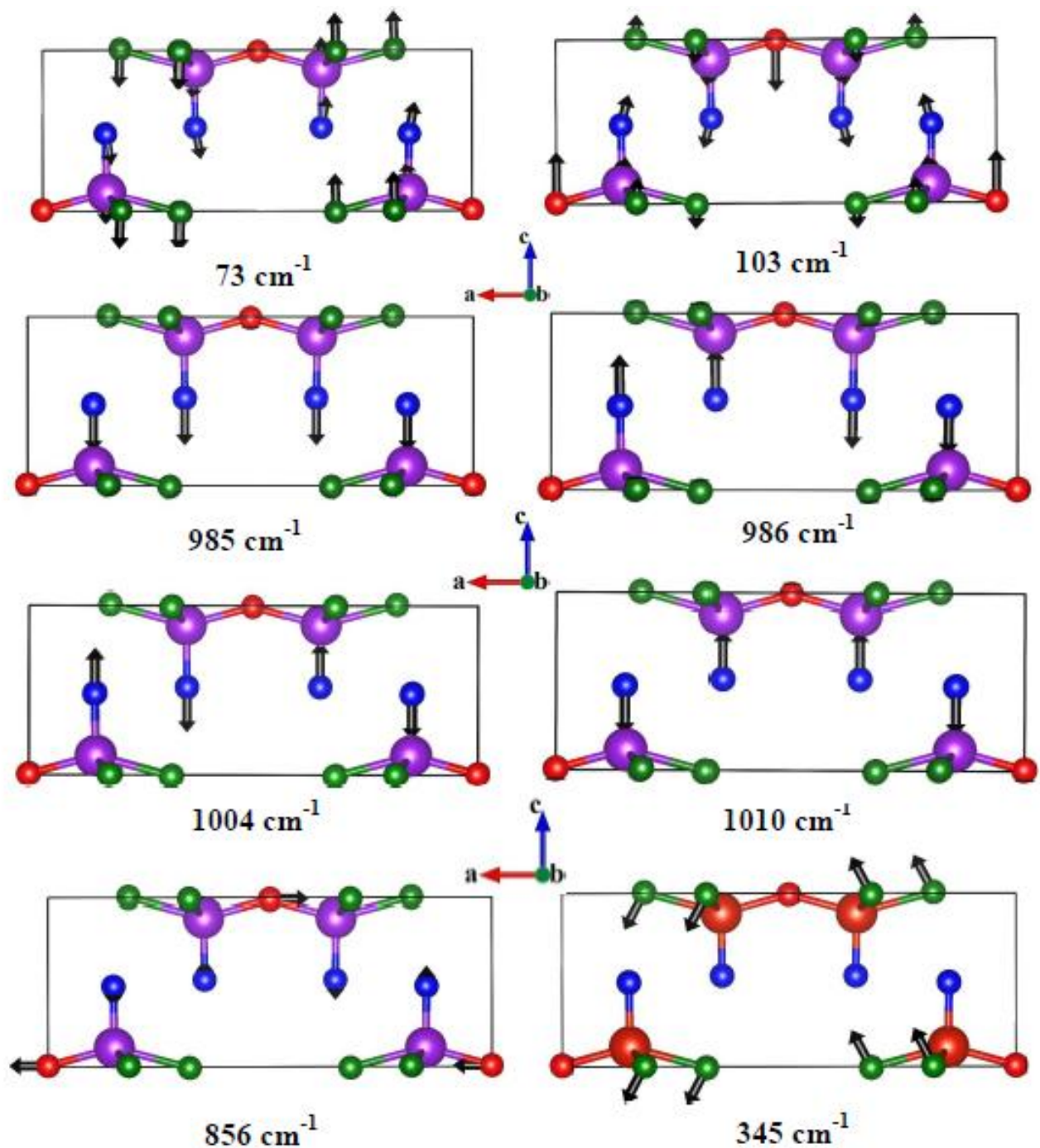

FIG. 7 (Color online) (a) The calculated anisotropic Grüneisen parameters. (b) The experimental and calculated lattice parameters as a function of temperature. (c) & (d) calculated linear thermal expansion coefficient as a function of phonon mode energy and temperature. (e) and (f) Eigenvectors of selected phonon modes that significantly contribute to the anomalous thermal expansion.

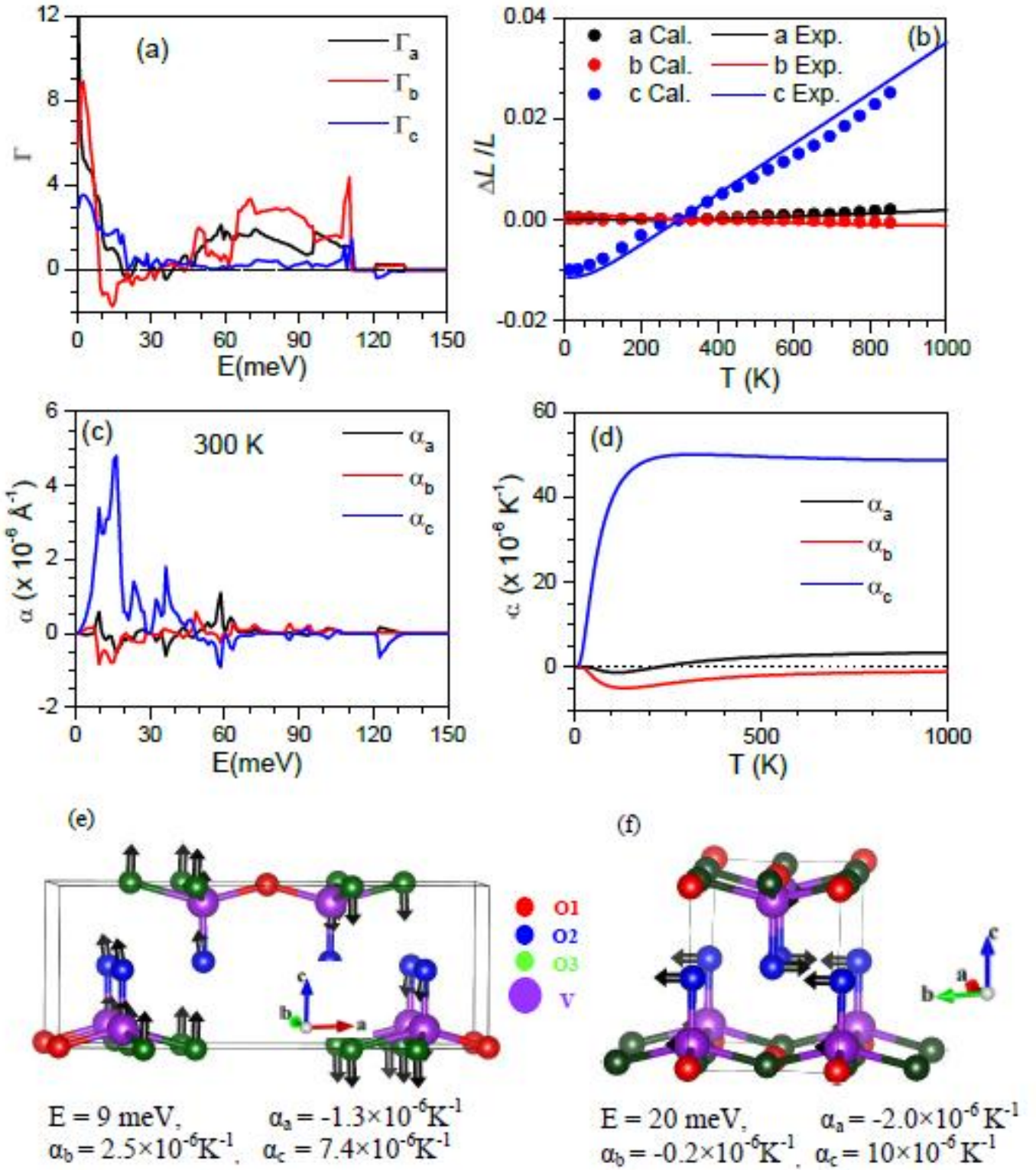

Fig 8. (Color online) Calculated and experimental[93] pressure dependence of lattice parameters, bond lengths and bond angles of orthorhombic α- $V_2O_5$. The calculations are shown in the pressure range of stability as observed experimentally[40, 93].

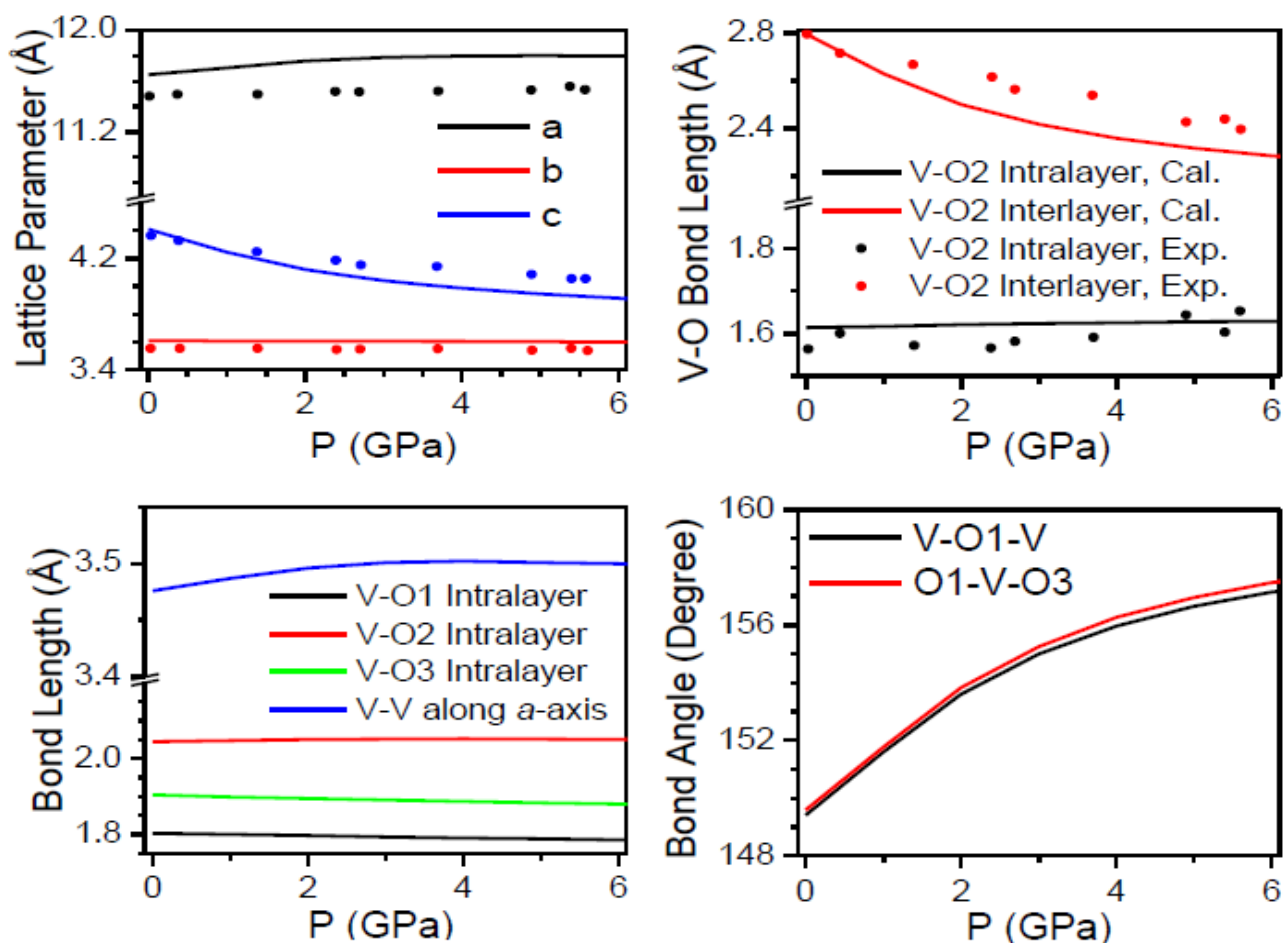

FIG 9. Calculated (0 GPa and 0K) phonon dispersion relation and phonon densities of states of various phases of $V_2O_5$. The dispersion curves are plotted through various high symmetry points of the Brillouin zone. The Bradley-Cracknell notation used for the high-symmetry points are given on the right.

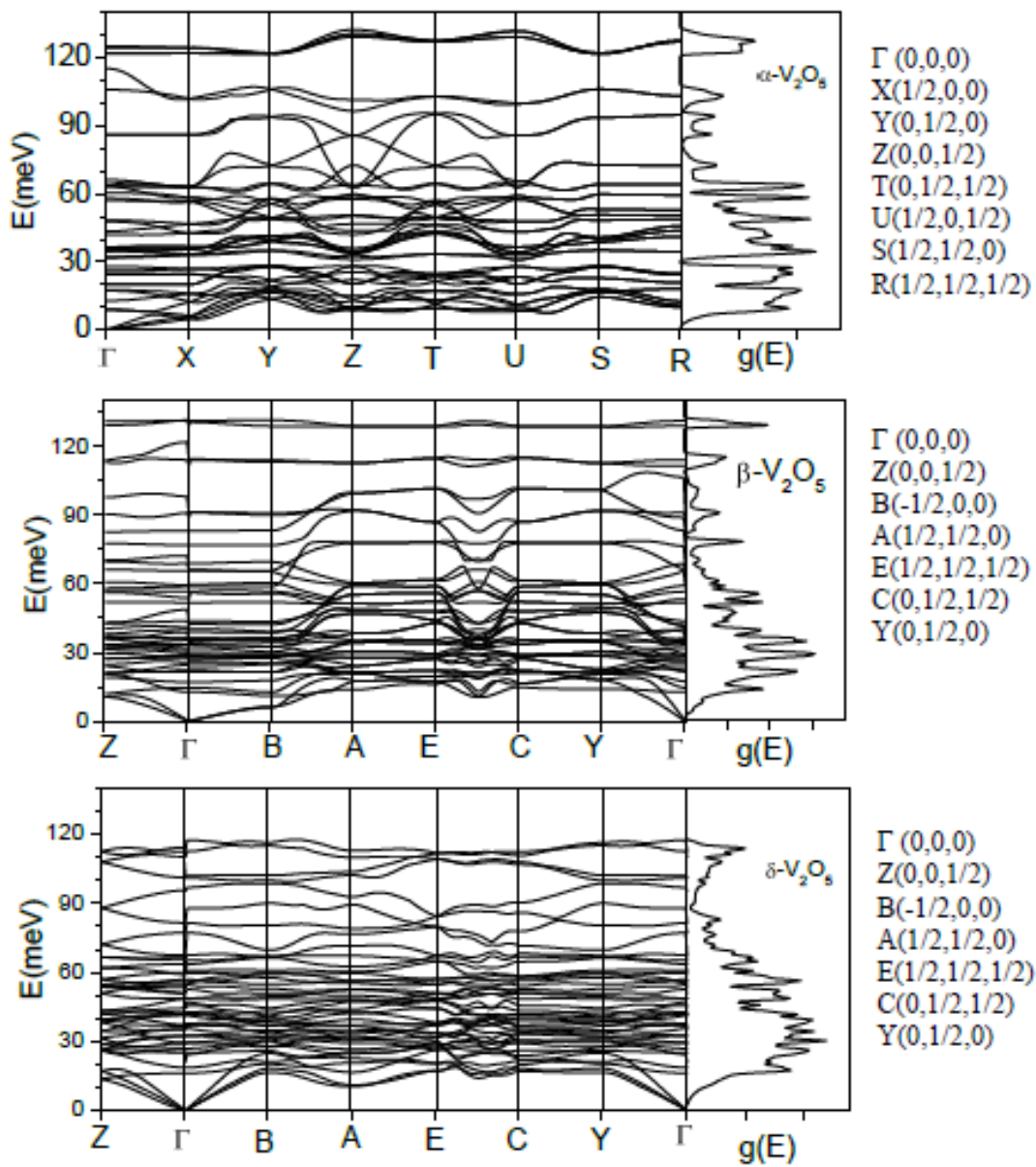

FIG.10 (Color online) Calculated Gibbs free energy per atom in various phases of $V_2O_5$ as a function of pressure at various temperatures.

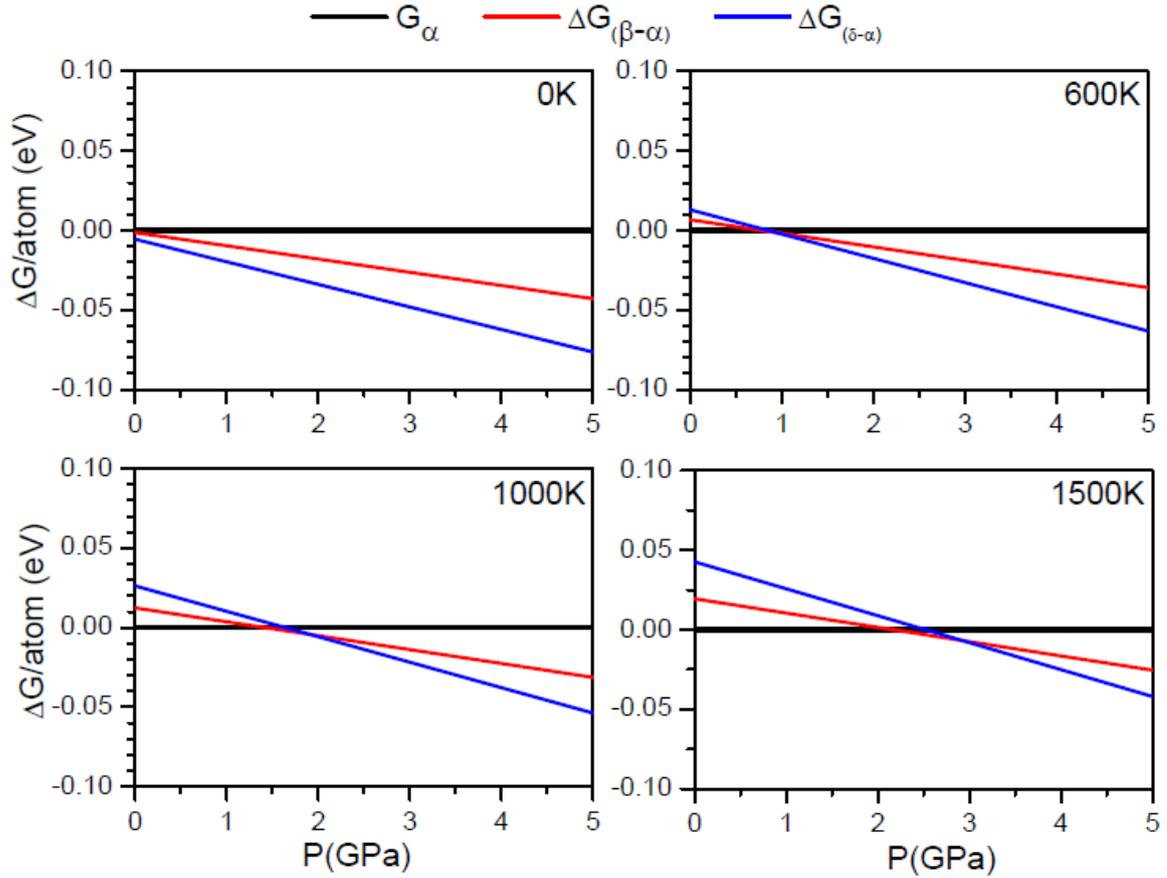

FIG.11 (Color online) Calculated and experimental[40] pressure/temperature phase diagram of $V_2O_5$. The experimental data points and calculated colored region are shown.

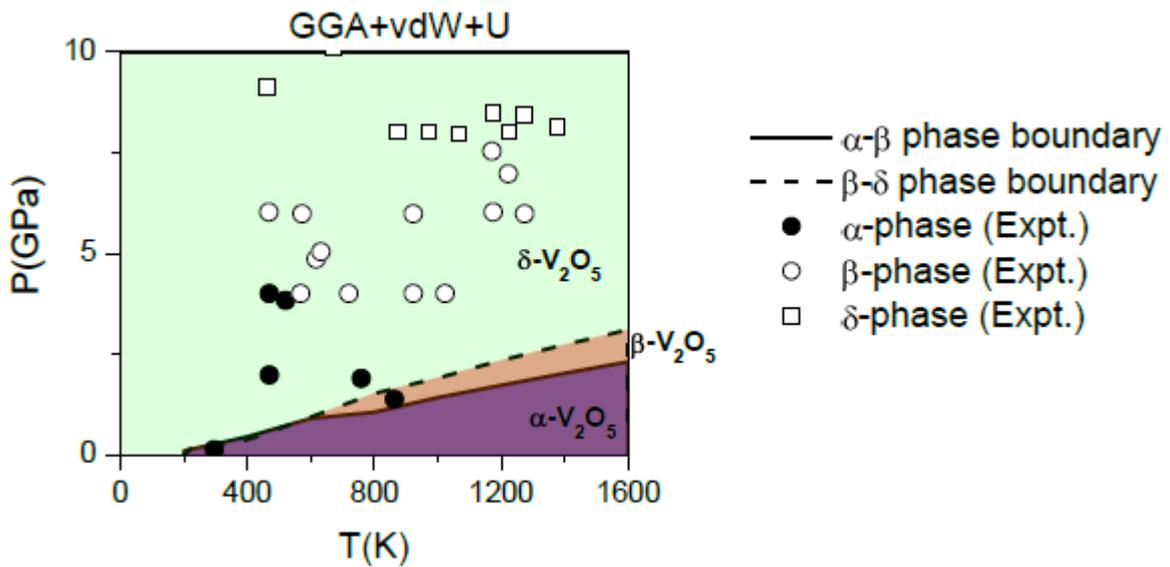

FIG.12 (Color online) Calculated volume per atom as a function of pressure for the α, β and δ- $V_2O_5$. The calculations are shown in the range of stability of various phases as observed experimentally[40, 93].

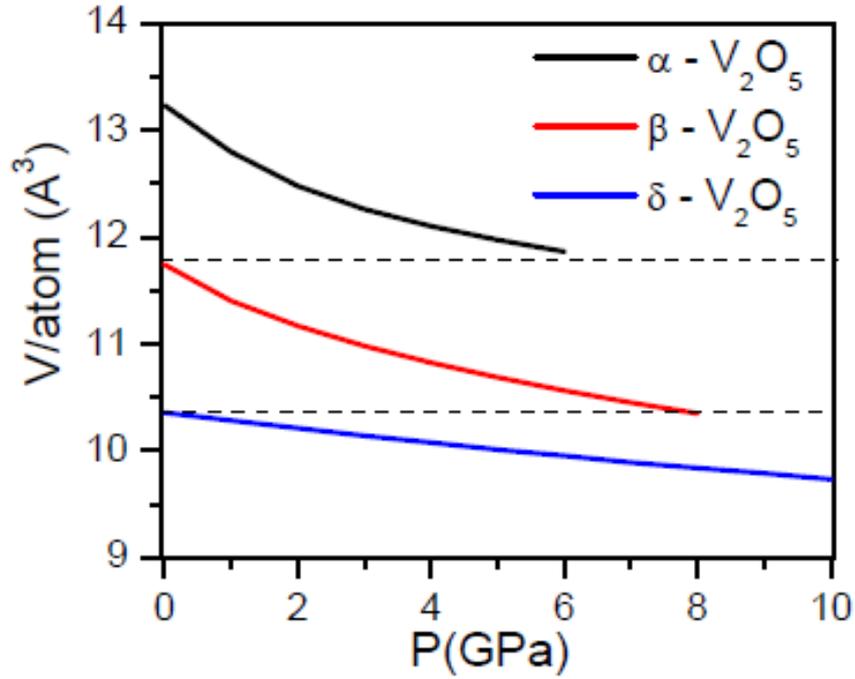

FIG.13 (Color online) Calculated elastic constant tensor as a function of pressure for α-$V_2O_5$. The calculations are shown in the pressure range of stability as observed experimentally[40],[93].

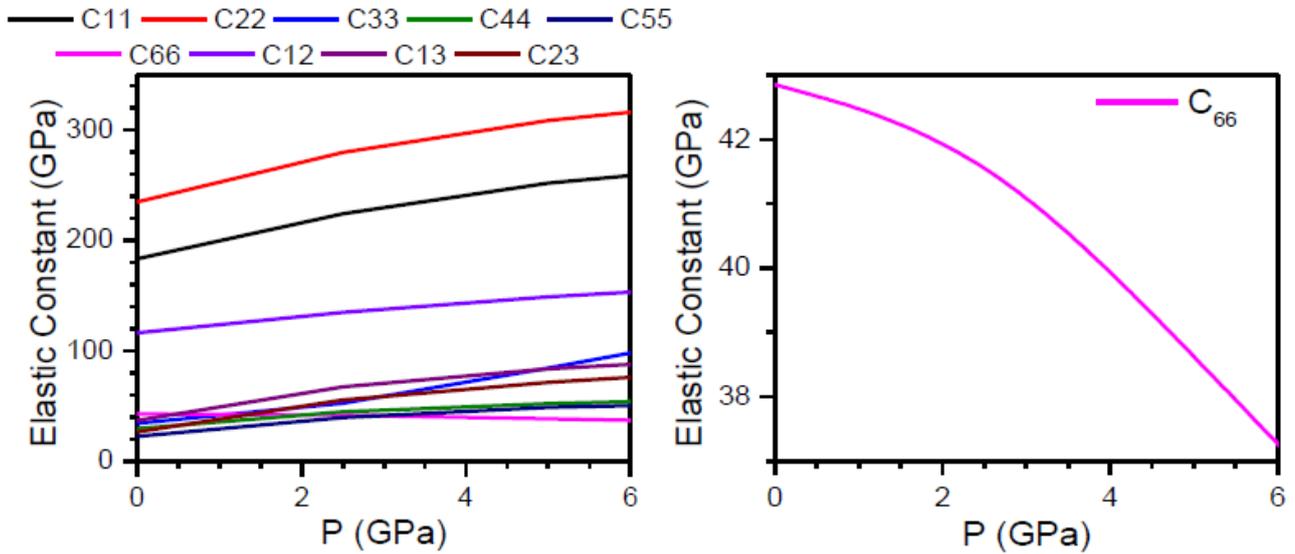